\documentclass[journal,twoside,web]{ieeecolor}
\usepackage[T1]{fontenc} 
\usepackage{generic}
\usepackage{cite}
\usepackage{amsmath,amssymb,amsfonts}
\usepackage{algorithmic}
\usepackage{graphicx}
\usepackage{multirow}
\usepackage{algorithm,algorithmic}
\usepackage{hyperref}
\hypersetup{hidelinks}
\usepackage{textcomp}
\usepackage{epstopdf}
\def\BibTeX{{\rm B\kern-.05em{\sc i\kern-.025em b}\kern-.08em
    T\kern-.1667em\lower.7ex\hbox{E}\kern-.125emX}}

\begin{document}
\title{Using Photoplethysmography to Detect Real-time Blood Pressure Changes with a Calibration-free Deep Learning Model}
\author{Jingyuan Hong, Manasi Nandi, Weiwei Jin, and Jordi Alastruey
\thanks{This work was supported by the Engineering and Physical Sciences Research Council Doctoral Training Partnership Grant [EP/T517963/1] and by the British Heart Foundation [PG/17/50/32903]. }
\thanks{Jingyuan Hong is with the Department of Biomedical Engineering, School of Biomedical Engineering and Imaging Sciences, King’s College London, King’s Health Partners, London SE1 7EU, U.K. (e-mail: jingyuan.hong@kcl.ac.uk). }
\thanks{Manasi Nandi is with the School of Cancer and Pharmaceutical Science, King’s College London, King’s Health Partners, London SE1 7EU, U.K. (e-mail: manasi.nandi@kcl.ac.uk).}
\thanks{Weiwei Jin is with the Department of Biomedical Engineering, School of Biomedical Engineering and Imaging Sciences, King’s College London, King’s Health Partners, London SE1 7EU, U.K. (e-mail: weiwei.jin@kcl.ac.uk).}
\thanks{Jordi Alastruey is with the Department of Biomedical Engineering, School of Biomedical Engineering and Imaging Sciences, King’s College London, King’s Health Partners, London SE1 7EU, U.K. (e-mail: jordi.alastruey-arimon@kcl.ac.uk).}}

\maketitle

\begin{abstract}
Blood pressure (BP) changes are linked to individual health status in both clinical and non-clinical settings. This study developed
a deep learning model to classify systolic (SBP), diastolic (DBP), and mean (MBP) BP changes using photoplethysmography (PPG) waveforms. 
Data from the Vital Signs Database (VitalDB) comprising 1,005 ICU patients with synchronized PPG and BP recordings was used. BP changes were categorized into three labels: Spike (increase above a threshold), Stable (change within a $\pm$ threshold), and Dip (decrease below a threshold). Four time-series classification models were studied: multi-layer perceptron, convolutional neural network, residual network, and Encoder. A subset of 500 patients was randomly selected for training and validation, ensuring a uniform distribution across BP change labels. Two test datasets were compiled: Test-I (n=500) with a uniform distribution selection process, and Test-II (n=5) without. The study also explored the impact of including second-deviation PPG (sdPPG) waveforms as additional input information. The Encoder model with a Softmax weighting process using both PPG and sdPPG waveforms achieved the highest detection accuracy—exceeding 71.3\% and 85.4\% in Test-I and Test-II, respectively, with thresholds of 30 mmHg for SBP, 15 mmHg for DBP, and 20 mmHg for MBP. Corresponding F1-scores were over 71.8\% and 88.5\%. These findings confirm that PPG waveforms are effective for real-time monitoring of BP changes in ICU settings and suggest potential for broader applications.

\end{abstract}

\begin{IEEEkeywords}
Photoplethysmography, Blood pressure monitoring, Deep learning classification model
\end{IEEEkeywords}

\section{Introduction}
\label{sec:introduction}
\IEEEPARstart{C}{hanges} in blood pressure (BP) are critical for understanding physiological conditions in both clinical and non-clinical settings. Real-time changes provide direct insights into the current health status of individuals. In a clinical setting, acute and severe changes in resting BP need for prompt alerts to facilitate timely medical intervention \cite{yu_expert_2023}. Furthermore, continuous BP monitoring allows for the evaluation of surgical risks \cite{berends_circulating_2021} and the tracking of postoperative complications \cite{luo_effect_2022}. In daily settings, monitoring BP changes during various activities aids in cardiovascular disease prevention, assessing sleep quality, and supervising exercise across diverse populations \cite{mchill_rapid_2022,serinel_diurnal_2019,wang_cardiovascular_2022}.

Three primary methods are employed to measure BP changes. Clinically, the most accurate involves invasive BP monitoring using catheters inserted into arteries, which allows for real-time BP waveform capture \cite{xiong_distal_2022}. Despite the precision of this method, it can cause considerable discomfort and pose health risks to patients \cite{lu_discrepancy_2023,alexander_chapter_2013,romagnoli_accuracy_2014}. For less invasive alternatives, clinicians use non-invasive continuous monitoring devices or perform multiple measurements with a cuff-based BP monitor \cite{meidert_techniques_2018,mancia_white-coat_2021}. While these methods reduce patient discomfort, their reliability and accuracy can be challenged \cite{mukkamala_evaluation_2021,chia_role_2022,panula_advances_2023}. In nonclinical environments, digital BP monitors are commonly used for home monitoring \cite{cuspidi_office_2020,mancia_short-_2021}. These devices measure BP at specific intervals to detect fluctuations. However, they are unable to provide accurate or continuous beat-to-beat BP readings, limiting their ability to capture sudden BP changes \cite{myers_limitations_2015,kinoshita_very_2022}.

An emerging alternative is the use of Photoplethysmography (PPG) signals. PPG, a technique that detects changes in peripheral blood volume, offers a novel method for predicting cardiovascular events \cite{charlton_wearable_2022}. Although the use of PPG signals to estimate absolute BP values has been explored for many years, developing models with high accuracy has been challenging, and most models require a moderate amount of the subject's baseline signal data for personalized calibration together with demographic information such as the subject's age \cite{ma_kd-informer_2023, liu_bigru-attention_2024}. However, rather than estimating absolute BP values, the categorical prediction of BP changes provides a promising pathway towards unobtrusive detection using continuous PPG signals based on the relationship between PPG and change of BP \cite{natarajan_photoplethysmography_2022}.

This study aims to detect BP changes from PPG signals using time-series classification deep learning models. Four models were trained and tested on 1,005 ICU patients from the Vital Signs Database (VitalDB), which includes time-aligned systolic (SBP) and diastolic (DBP) BP values together with continuous PPG signals for each patient. BP changes were categorized into significant increases, normal range changes, and significant decreases, with thresholds set from ±5 mmHg to ±45 mmHg in ±5 mmHg increments. Additional analyses were conducted using the Encoder model, which outperformed the other three models studied.

\section{Methodology}

\subsection{Data}
VitalDB contains continuous PPG and BP signals both with a sampling rate of 125 Hz for 2,938 ICU patients \cite{wang_pulsedb_2023}. In this dataset, finger PPG and left radial BP waveforms were measured by patient monitors (Tram-Rac 4A, GE healthcare) over periods ranging from 10 seconds to 10 hours \cite{lee_vitaldb_2022}. For convenience, PPG and BP for each patient are provided split into non-overlap 10-second segments. In this study, only patients with recording times longer than 30 minutes were included, resulting in 2,131 patients, with characteristics shown in Table \ref{tab1}. 

\begin{table*}
\centering
\caption{Characteristics of the VitalDB cohorts used in this study}
\label{tab1}
\setlength{\tabcolsep}{3pt}
\begin{tabular}{p{150pt}p{60pt}p{90pt}p{60pt}p{60pt}}
\hline
Index& 
VitalDB&Training \& Validation&Test-I&Test-II\\
\hline
Number of patients& 2,131 & 500& 500& 5
\\
10-second segments per patient& 662$\pm$384 & 412$\pm$200 & 429$\pm$212 & 399$\pm$87
\\
Recording duration, minutes& 110.3$\pm$64.0 & 68.6$\pm$33.3 & 71.5$\pm$35.3 & 66.5$\pm$14.5
\\
BP change segments before sampling& 6.2$\times10^8$& 5.2$\times10^7$ & 5.7$\times10^7$ & 4.0$\times10^5$
\\
BP change segments after sampling& N/A & 2.7$\times10^6$ & 1.4$\times10^5$ & N/A
\\
Age, yr& 58.9$\pm$14.9 & 58.6$\pm$14.8 & 58.5$\pm$15.2 & 64.2$\pm$14.0
\\
Sex, Female, \%& 44.3 & 42.8 & 41.8 & 80.0
\\
Height, m& 1.62$\pm$0.09 & 1.63$\pm$0.09 & 1.63$\pm$0.09 & 1.58$\pm$0.06
\\
Weight, kg& 61.0$\pm$11.8 & 60.9$\pm$12.0 & 61.5$\pm$11.8 & 60.5$\pm$10.8
\\
\hline
\multicolumn{5}{p{450pt}}{Values are Mean$\pm$SD or n. Test-I: data from 500 patients with an uniform distribution of the three classification labels, mirroring the distribution in the training and validation datasets. Test-II: comprehensive BP change data from a smaller cohort of 5 patients. N/A: not applicable.}\\
\end{tabular}
\end{table*}

\subsection{Calculation of BP changes}
Changes in three types of BP were calculated. Single SBP and DBP values were calculated as the mean SBP and DBP from each 10-second BP recording. The MBP value was then calculated from the SBP and DBP values using the formula (SBP + 2DBP)/3 \cite{sainas_mean_2016}. For each patient, changes in each of the three BP types between two time points, $\Delta BP$, were calculated as:
\begin{equation}
\begin{split}
    \Delta BP = BP_{i+j}-BP_{i}\\
    i= [1,2,3,...,N-1],\\
    j=[1,2,3,...,N-i]  
\end{split}
\end{equation}
where \textit{i} is the index of an initial BP reading, \textit{j} is the number of subsequent readings after \textit{i} at which the BP was measured, and \textit{N} is the total number of BP readings. This formula captures BP changes over varying time intervals for each patient. 

\begin{figure}[!t]
\centerline{\includegraphics[width=200pt]{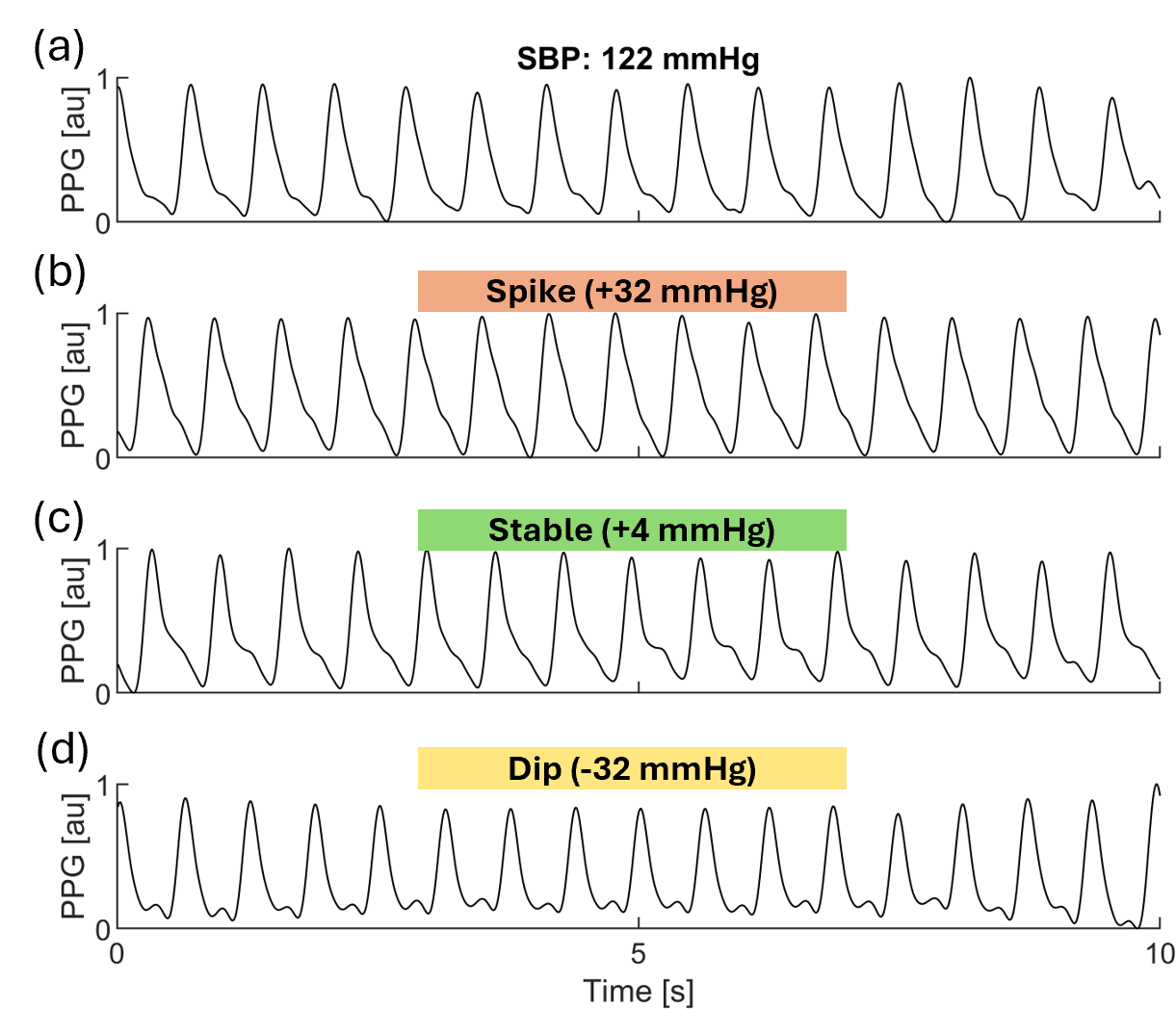}}
\caption{Examples of 10-second PPG signals with corresponding systolic blood pressure (SBP) values: (a) initial SBP reading, (b) Spike change in SBP, (c) Stable SBP, and (d) Dip change in SBP.}
\label{fig1}
\end{figure}

BP changes were categorized into three labels: "Spike", "Stable", and "Dip". These labels were defined as a significant increase, maintenance within a normal range, and a significant decrease, respectively. Initial thresholds for these changes were set at 30 mmHg for SBP, 15 mmHg for DBP, and 20 mmHg for MBP based on the 75\% confidence interval of calculated changes in SBP, DBP, and MBP distributions. Increases greater than 30 mmHg for SBP, 15 mmHg for DBP, or 20 mmHg for MBP were classified as Spike. Changes within $\pm$30 mmHg for SBP, $\pm$15 mmHg for DBP, or $\pm$20 mmHg for MBP were classified as Stable. Decreases greater than 30 mmHg for SBP, 15 mmHg for DBP, or 20 mmHg for MBP were classified as Dip. These thresholds were adjusted for further analysis to a range from 5 to 45 mmHg for SBP, 5 to 35 mmHg for DBP, and 5 to 40 mmHg for MBP, in 5 mmHg increments each. The maximum thresholds for the three BP types differ due to insufficient data, with the frequency of changes exceeding 45 mmHg in DBP and MBP being lower than in SBP. An example of PPG signals with different SBP labels is shown in Fig. \ref{fig1}.

\subsection{Classification models}
Detecting BP change categories from PPG waveforms is a typical time-series classification task, for which four deep learning models were used in this study \cite{fawaz_deep_2019}.

\begin{figure}[!t]
\centerline{\includegraphics[width=\columnwidth]{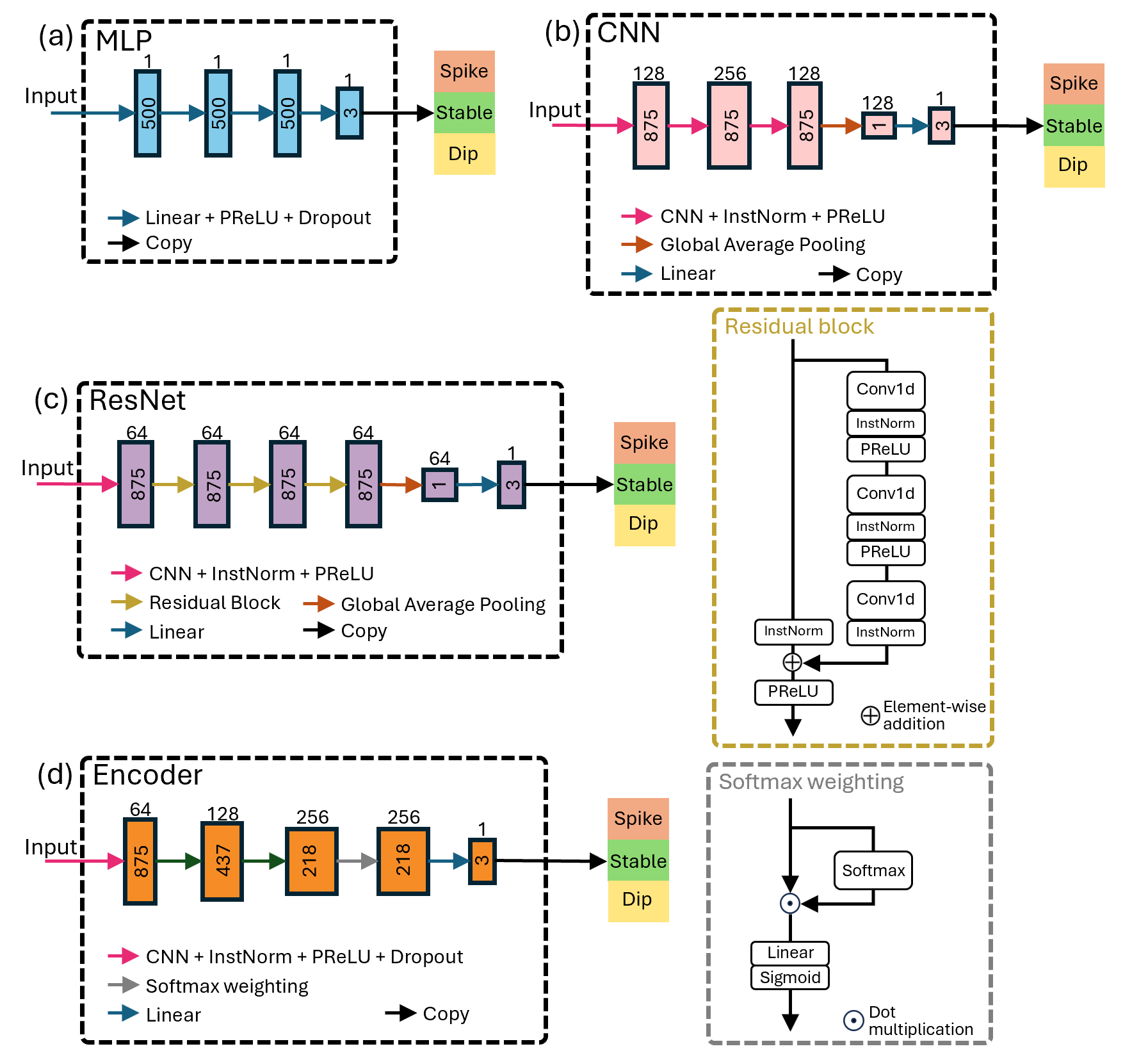}}
\caption{Architectures of the four classification models: (a) multi-layer perceptron (MLP), (b) convolutional neural network (CNN), (c) residual network (ResNet) with detailed residual block, and (d) Encoder with detailed Softmax weighting process.}
\label{fig2}
\end{figure}

Multi-layer perceptron (MLP) is a basic deep learning architecture. In this study, the MLP model comprised four fully connected layers (Fig. \ref{fig2}a). The first layer flattened all input channels, while the following layers each consisted of 500 neurons with a parametric rectified linear unit (PReLU) activation function and a dropout layer. The final layer classified the outputs into three categories.

Convolutional neural network (CNN) models are effective for extracting deep features from time-series data. In this study, the CNN model included three convolution blocks (Fig. \ref{fig2}b), each followed by an instance normalization layer and a PReLU activation function, a global average pooling layer, and a final fully connected layer that classified into three categories.

Residual network (ResNet) modifies a traditional CNN  model by adding shortcut residual connections between convolutional layers, which helps avoid the vanishing gradient problem and improves the ability of the network to learn from deep architectures. In this study, the ResNet model consisted of three sequential residual blocks (Fig. \ref{fig2}c), each containing three CNN blocks (a conventional layer, an instance normalization layer, and a PReLU activation function), followed by a global average pooling layer and a final fully connected layer.

Encoder (from Transformer architecture) combines a CNN model with an attention mechanism. In this study, the Encoder model contained three CNN blocks (each with a conventional layer, an instance normalization layer, a PReLU activation function, and a dropout layer), an attention mechanism, and a final fully connected layer (Fig. \ref{fig2}d). The attention mechanism used a Softmax weighting process, starting by applying the Softmax function to the output of the last CNN block to generate a set of normalized weights. These weights then scaled the corresponding features, emphasizing informative parts and diminishing less relevant ones. The weighted features were summed to produce an attention-augmented output, helping the model focus on specific and relevant parts of the time-series inputs.

The architecture details and optimization hyperparameters of these four models are summarized in Table \ref{tab2}. All models in this study were built using the Pytorch framework and trained on an Intel Xeon W-2195 CPU (2.3 GHz) with an NVDIA Tesla V100 GPU.

\begin{table}
\centering
\caption{Summary of hyperparameters for the deep learning models}
\label{tab2}
\setlength{\tabcolsep}{3pt}
\begin{tabular}{p{48pt}p{46pt}p{46pt}p{46pt}p{46pt}}
\hline
 & MLP& CNN& ResNet& Encoder
\\
\hline
\multicolumn{5}{c}{Architecture}
\\
\hline
Layers& 4& 4& 11&5
\\
Convolutional & 0& 3& 10&3
\\
Normalization& None& Instance& Instance&Instance
\\
Pooling& None& Average& Average&Max
\\
Activate& PReLU& PReLU& PReLU&PReLU
\\
\hline
\multicolumn{5}{c}{Optimization}
\\
\hline
Algorithm & Adam& Adam& Adam&Adam
\\
Loss function & Cross Entropy& Cross Entropy& Cross Entropy&Cross Entropy
\\

Epochs & 1024&1024 & 1024&1024
\\
Batches & 500&500 & 500&500
\\
Learning rate& 0.001& 0.001& 0.001&0.0001\\
\hline
\end{tabular}
\end{table}

\subsection{Input data types}
\label{app:inputs}
Given the strong correlation between the second-derivative PPG (sdPPG) waveform and BP \cite{shin_feasibility_2017}, this study compared three input combinations to evaluate the impact of incorporating the sdPPG waveform in model training. Each combination was applied to detect changes in SBP, DBP, and MBP.

The first input type included only the PPG waveform (PPG-waveform). This input type provided PPG waveforms at time steps \textit{i} and \textit{i}+\textit{j} (Fig. \ref{fig3}b) and served as the basic input for evaluating the classification performance of the proposed four models.

The second input type integrated five features were derived from the sdPPG waveform and added them to the PPG-waveform input (Waveform-feature). This input type included the PPG waveform and the extracted sdPPG features at time steps \textit{i} and \textit{i}+\textit{j} (Fig. \ref{fig3}b). The definition of these five features is shown in Table \ref{tab3} and their selection process is detailed in Appendix \ref{app:features}.

The third input type included the PPG waveform along with the sdPPG waveform (PPG-sdPPG-waveform). Both the PPG and sdPPG waveforms were provided at time steps \textit{i} and \textit{i}+\textit{j} (Fig. \ref{fig3}b). Min-max normalization was applied to the sdPPG waveform to match its amplitude range with the primary PPG input.

To standardize all PPG waveforms, the original 10-second PPG recordings were truncated to 7 seconds to ensure each signal began with a complete cardiac cycle. From this 7-second baseline, the impact of input length on detection performance was evaluated by testing the Encoder model with PPG-sdPPG-waveform input at 3, 5, and 7-second lengths.

Besides the PPG input, the BP value at time step \textit{i} was also fed into the model as an extra input (Fig. \ref{fig3}c). An ablation study was conducted for the Encoder model with the PPG-sdPPG-waveform input type to assess the necessity of using initial BP values as an extra input for training, and quantifying the effects of removing the initial BP on classification outcomes.

\begin{table}
\centering
\caption{PPG features used in the Waveform-feature input type}
\label{tab3}
\setlength{\tabcolsep}{3pt}
\begin{tabular}{p{120pt}p{120pt}}
\hline
Index& 
Formula 
 \\
\hline
$b/a $& 
$\frac{b}{a} $
\\
&  \\
$slope_{b-c}$& 
$\frac{b-c}{T_b-T_c}$
\\
&  \\
$slope_{b-d}$& 
$\frac{b-d}{T_b-T_d}$ 
\\
&  \\
$AGI$& 
$\frac{b-c-d-e}{a}$ 
 \\
 &  \\
$AGI_{mod}$& 
$\frac{b-c-d}{a}$
 \\
  &  \\
\hline
\multicolumn{2}{p{250pt}}{Variable definition: $a$, peak of early systolic positive wave of the second-derivative PPG (sdPPG); $b$, valley of early systolic negative wave of the sdPPG; $c$, peak of late systolic re-increasing wave of the sdPPG; $d$, valley of late systolic re-decreasing wave of the sdPPG; $e$, peak of early diastolic positive wave of the sdPPG; $T_b$, time at point b; $T_d$, time at point c; $T_d$, time at point d.}  \\
\end{tabular}
\end{table}

\subsection{Training, validation, and test}
\begin{figure}[!t]
\centerline{\includegraphics[width=\columnwidth]{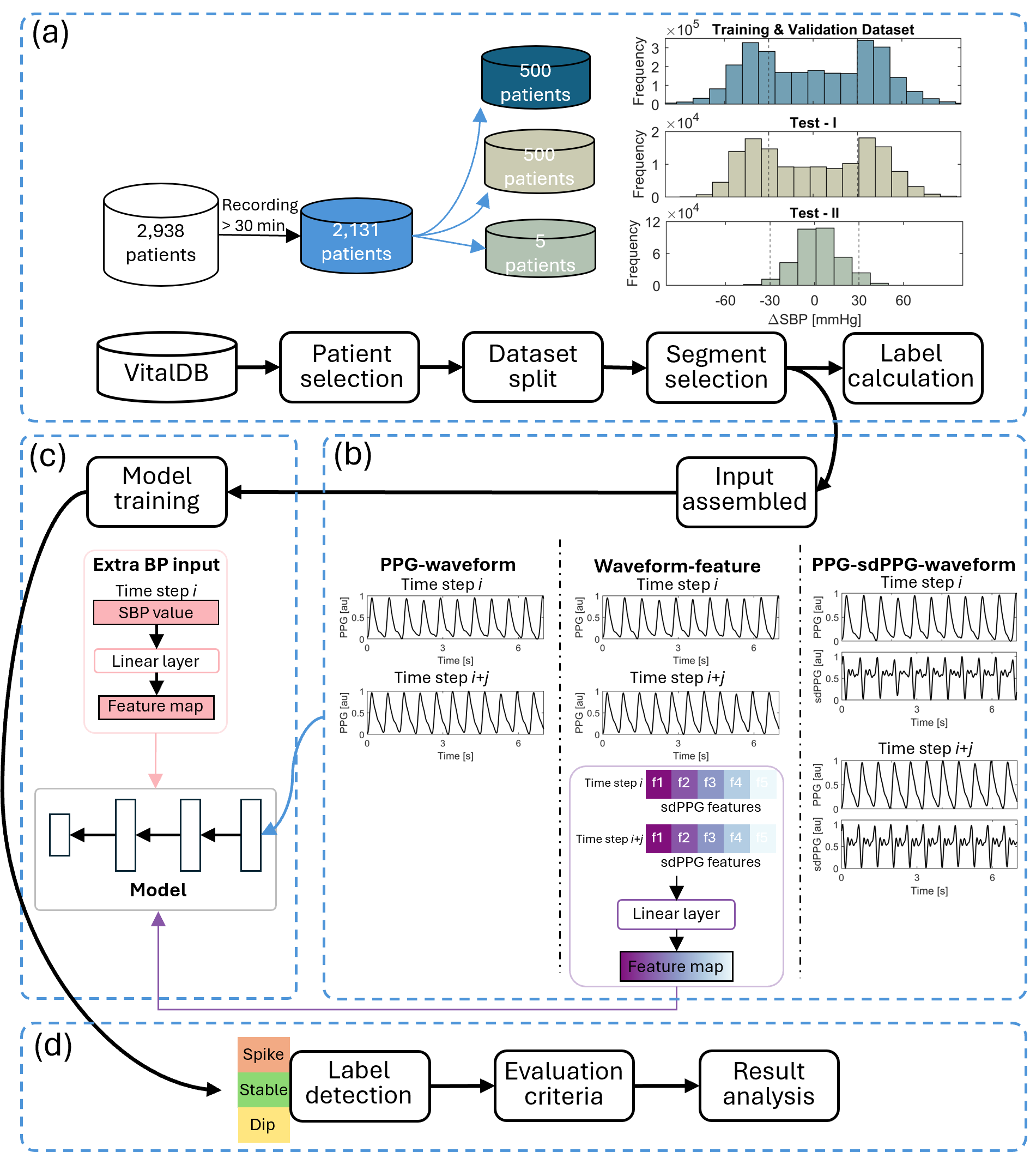}}
\caption{Flow chart of the overall experimental design using the example of detecting changes in SBP labels. (a) VitalDB was split into one training and validation dataset (n=500), a Test-I dataset (n=500) with a selected segments to achieve a similar  distribution of changes in SBP ($\Delta SBP$), and a Test-II dataset (n=5) containing all segments. (b) Three different input types were considered: (1) PPG-waveform input: time step \textit{i} and time step \textit{i}+\textit{j} PPG waveforms; (2) Waveform-feature input: PPG-waveform input plus 5 features extracted from time step \textit{i} and time step \textit{i}+\textit{j} second derivative PPG (sdPPG) waveforms; (3) PPG-sdPPG-waveform: time step \textit{i} and time step \textit{i}+\textit{j} PPG waveforms and sdPPG waveforms. (c) Model training process with extra initial BP input. (d) Detection of BP changes categorized into three labels: Spike, Stable, and Dip, followed by result evaluation using accuracy and F1-score metrics.}
\label{fig3}
\end{figure}

The training, validation and test pipeline for all models and input types is illustrated in Fig. \ref{fig3}. To manage resource constraints and avoid over-fitting, we randomly selected 500 patients from the total of 2,131 patients (Fig. \ref{fig3}a). A sampling process was then implemented to ensure a uniform distribution by selecting an equal number of segments across the three categories (Table \ref{tab1}): Spike (900,000 segments), Stable (900,000 segments), and Dip (900,000 segments). This selection aimed to train the model without bias towards any of the three output categories. 

Two test datasets were created as follows (Fig. \ref{fig3}a): Test dataset I (Test-I) involved selecting 500 patients from the total of 2,131 patients, excluding those in the training and validation dataset. From all test segments, 144,500 segments were selected to match the uniform distribution sampled in the training and validation dataset (Table \ref{tab1}). Test dataset II (Test-II) involved selecting 5 patients from the total of 2,131 patients, excluding those from the training and validation dataset and Test-I dataset, and using all their segments (Table \ref{tab1}). The characteristics of these datasets are shown in Table \ref{tab1}.

For the training and validation dataset, 80\% was allocated for training and 20\% for validation, with a five-fold cross validation method applied to the training process of all models. The model with the best performance was retained for further analysis.

Each model was trained using the PPG-waveform input type as described in Section \ref{app:inputs} (Fig. \ref{fig3}b). Furthermore, the Encoder model (best performer) was trained and tested for the three input types described in Section \ref{app:inputs} (Fig. \ref{fig3}b). For each input type, the Encoder was trained with or without the initial BP value at time step \textit{i}. This value was incorporated into the model as supplementary information, using a linear layer to concatenate it with the feature map from the last layer before inputting it into each layer of the model (Fig. \ref{fig3}c). The model produced three output categories that were evaluated for detection results using accuracy and F1-score metrics (Fig. \ref{fig3}d), as described next.

\subsection{Evaluation metrics}
Cross-entropy was employed as the loss function to transform the true label probability predicted by the model into a loss value. This approach aimed to minimize the loss value, ensuring that the predictive probability distribution of the model closely approximates the true label distribution,
\begin{equation}
\mathcal{L}=H(p,q)=\sum^{C}_{x=1}p(x)\log q(x)
\end{equation}
where $\mathcal{L}$ is the value of the cross-entropy loss function, $H$ denotes the cross-entropy used to measure the discrepancy between the two probability distributions, $x$ is an index representing different categories, $C$ represents the total number of categories, $p(x)$ is the true label probability for class $x$, and $q(x)$ is the predicted probability for class $x$. 

Two metrics were used for evaluating the classification results of the proposed models. Accuracy measured the proportion of samples that were correctly predicted by the model relative to the total number of samples. F1-score served as a balanced indicator to assess both the accuracy and robustness of the model, considering both precision and recall. The mathematical definitions of these metrics are as follows:

\begin{equation}
Accuracy=\frac{TP+TN}{TP+TN+FP+FN}
\end{equation}
\begin{equation}
Precision = \frac{TP}{TP+FP}
\end{equation}
\begin{equation}
Recall = \frac{TP}{TP+FN}
\end{equation}
\begin{equation}
F1\mbox{-}score = 2\times\frac{Precision\times Recall}{Precision+Recall}
\end{equation}
\begin{equation}
F1\mbox{-}score_{total} = \frac{1}{C}\sum^{C}F1\mbox{-}score
\end{equation}
where \textit{TP}, \textit{TN}, \textit{FP}, and \textit{FN} represent the counts of true positives, true negatives, false positives, and false negatives, respectively.

\section{Results}

\subsection{Model evaluation}
The Encoder model achieved the highest accuracy and F1-score values across the three BP types for the two test datasets studied (Table \ref{tab4}). Changes in MBP were detected more accurately compared to changes in SBP and DBP in both Test-I and Test-II datasets. All models and pressure types showed improved performance on the Test-II dataset compared to the Test-I dataset. Overall, the Encoder model considerably improved detection performance. Compared to the MLP model on the Test-I dataset, BP changes detection accuracy increased by $\ge$5.2\% and the F1-score by $\ge$5.1\% for all pressure types. On the Test-II dataset, these increases were $\ge$3.7\% in accuracy and $\ge$2.6\% in F1-score.

\begin{table}
\centering
\caption{Classification performance of the proposed deep learning models with PPG-waveform input}
\label{tab4}
\setlength{\tabcolsep}{3pt}
\begin{tabular}{p{30pt}p{25pt}p{25pt}p{25pt}p{25pt}p{25pt}p{25pt}}
\hline
Model &\multicolumn{3}{c}{Accuracy (\%)}& \multicolumn{3}{c}{F1-score (\%)}
\\
\hline
& SBP& DBP& MBP& SBP& DBP& MBP
\\
\hline
\multicolumn{7}{c}{Test-I}
\\
\hline
MLP & 62.6& 68.6& 70.7& 62.4& 68.9&71.3
\\
CNN & 68.5& 70.7& 72.1& 69.0& 71.0&72.6
\\
ResNet & 69.2& 74.3& 73.9& 69.7& 74.6&74.5
\\
Encoder & \textbf{70.2}& \textbf{75.1}& \textbf{75.9}& \textbf{70.4}& \textbf{75.5}&\textbf{76.4}
\\
\hline
\multicolumn{7}{c}{Test-II}
\\
\hline
MLP & 77.5& 82.8& 87.2& 83.1& 86.4& 89.6
\\
CNN  & 81.3& 82.3& 88.0& 85.8& 86.2&90.3
\\
ResNet & 82.4& 81.6& 87.7& 86.5& 85.9&90.2
\\
Encoder & \textbf{82.8}& \textbf{86.3}& \textbf{90.9}& \textbf{86.7}& \textbf{89.0}&\textbf{92.2}
\\
\hline
\end{tabular}
\end{table}

\subsection{Input evaluation}
The classification performance of the Encoder model for both test datasets improved when the sdPPG waveform was used as an additional input (PPG-sdPPG-waveform), compared to using only the basic PPG waveform (PPG-waveform) or the PPG waveform combined with five sdPPG features (Waveform-feature). The PPG-sdPPG-waveform input type yielded the highest accuracy and F1-score for detecting changes in all BP types on Test-I and in SBP and DBP on Test-II. Consequently, the PPG-sdPPG-waveform input was selected for subsequent analysis due to its superior performance.

\begin{figure}[!t]
\centerline{\includegraphics[width=0.8\columnwidth]{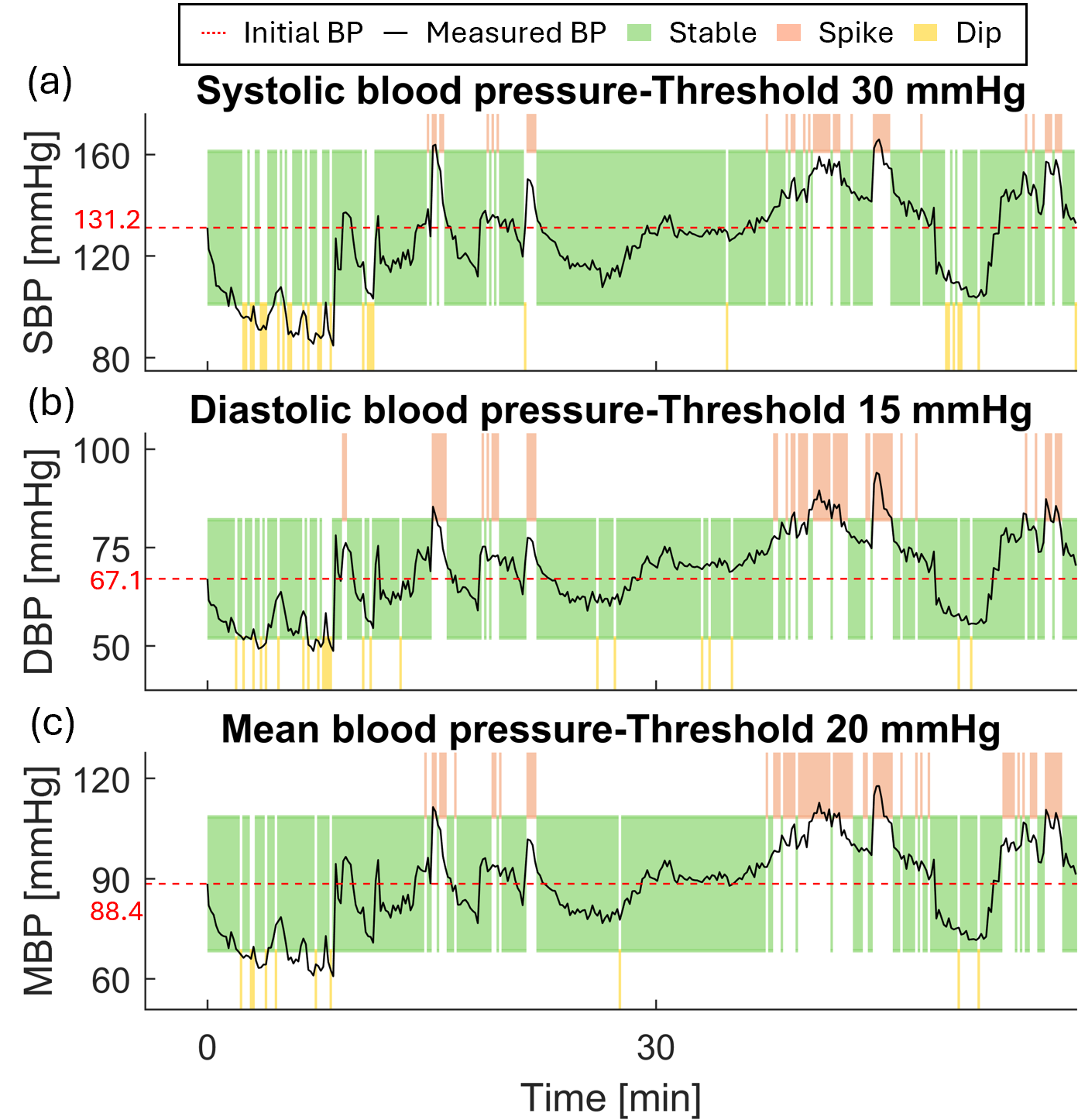}}
\caption{Label detection results from a patient in the Test-II dataset using the Encoder model with the PPG-sdPPG-waveform input type. Reference (a) systolic (SBP), (b) diastolic (DBP), and (c) mean (MBP) blood pressures (black lines), and corresponding detected Spike (red), Stable (green), and Dip (yellow) labels with respect to initial pressures calculated at time zero (dashed red lines). Note the different threshold values and range of y-axis values for each BP type.}
\label{fig4}
\end{figure}

Using the initial thresholds for changes in SBP, DBP and MBP based on the 75\% confidence intervals, most variations in the reference measured BP when setting the initial BP at time step zero fell within the ‘Stable’ range (green area). This was the most frequent classification label detected by the Encoder model with the PPG-sdPPG-waveform input on the Test-II dataset (Fig. \ref{fig4}). The model accurately detected the Spike and Dip labels when some portion of the reference BP sudden increased or decreased beyond the thresholds for all BP types. However, it occasionally mislabeled changes, especially when BP fluctuations were within the set thresholds. Fig. \ref{fig4} shows the results for one of the five patients in Test-II dataset, and Supplementary Material Figs. \ref{fig_sup1} to \ref{fig_sup5} show similar correspondences between reference values and detected BP changes labels for the remaining four patients at the same thresholds. All patients showed similar trajectories for the three BP reference values.

A decrease in the length of the PPG input for the Encoder model, with a PPG-sdPPG-waveform input type, reduced the detection accuracy and F1-score across all BP types on Test-I by less than 1.1\% and 3.2\%. However, these metrics increased up to 1.5\% and 0.8\%, respectively, on Test-II (Table \ref{tab7}). The ablation of the initial BP value for the Encoder model with the PPG-sdPPG-waveform input type decreased the classification accuracy and F1-score across all BP types by less than 0.9\% and 1.0\%, respectively, for the Test-I dataset, and 1.5\% and 1.3\%, respectively, for the Test-II dataset (Table \ref{tab6}).

\begin{table}
\centering
\caption{Classification performance of Encoder model with different input types}
\label{tab5}
\setlength{\tabcolsep}{3pt}
\begin{tabular}{p{75pt}p{18pt}p{18pt}p{18pt}p{18pt}p{18pt}p{18pt}}
\hline
Input types&\multicolumn{3}{c}{Accuracy (\%)}& \multicolumn{3}{c}{F1-score (\%)}
\\
\hline
& SBP& DBP& MBP& SBP& DBP& MBP
\\
\hline
\multicolumn{7}{c}{Test-I}
\\
\hline
PPG-waveform & 70.2& 75.1& 75.9& 70.4& 75.5&76.4
\\
Waveform-feature & 70.3& 74.2& 75.3& 70.6& 74.6&75.9
\\
PPG-sdPPG-waveform & \textbf{71.3}& \textbf{76.4}& \textbf{76.2}& \textbf{71.8}& \textbf{76.9}&\textbf{76.7}
\\
\hline
\multicolumn{7}{c}{Test-II}
\\
\hline
PPG-waveform & 82.8& 86.3& \textbf{90.9}& 86.7& 89.0& \textbf{92.2}
\\
Waveform-feature & 82.7& 85.4& 90.3& 86.6& 88.4& 91.9
\\
PPG-sdPPG-waveform & \textbf{85.4}& \textbf{87.3}& 88.8& \textbf{88.5}& \textbf{89.7}&90.8
\\
\hline
\end{tabular}
\end{table}

\begin{table}
\centering
\caption{Classification performance of Encoder model with varying signal length of the PPG-sdPPG-waveform input}
\label{tab7}
\setlength{\tabcolsep}{3pt}
\begin{tabular}{p{70pt}p{15pt}p{15pt}p{15pt}p{15pt}p{15pt}p{15pt}}
\hline
Length (seconds) &\multicolumn{3}{c}{Accuracy (\%)}& \multicolumn{3}{c}{F1-score (\%)}
\\
\hline
& SBP& DBP& MBP& SBP& DBP& MBP
\\
\hline
\multicolumn{7}{c}{Test-I}
\\
\hline
\multicolumn{1}{c}{3}  & 70.6& 73.3& 76.1& 71.0& 73.7& \textbf{76.7}\\
\multicolumn{1}{c}{5}  & 70.7& 76.1& 75.0& 71.3& 76.6& 75.6\\
\multicolumn{1}{c}{7} & \textbf{71.3}& \textbf{76.4}& \textbf{76.2}& \textbf{71.8}& \textbf{76.9}&\textbf{76.7}
\\
\hline
\multicolumn{7}{c}{Test-II}
\\
\hline
\multicolumn{1}{c}{3} & 84.4& \textbf{87.6}& 88.6& 87.8& \textbf{89.8}& 90.8\\
\multicolumn{1}{c}{5} & \textbf{86.9}& 85.1& \textbf{89.3}&\textbf{89.3}& 88.4& \textbf{91.1}\\
\multicolumn{1}{c}{7} & 85.4& 87.3& 88.8& 88.5& 89.7&90.8
\\
\hline
\end{tabular}
\end{table}

\begin{table}
\centering
\caption{Classification performance of Encoder model with the PPG-sdPPG-waveform input type, with and without the initial BP values}
\label{tab6}
\setlength{\tabcolsep}{3pt}
\begin{tabular}{p{93pt}p{15pt}p{15pt}p{15pt}p{15pt}p{15pt}p{15pt}}
\hline
Initial BP value &\multicolumn{3}{c}{Accuracy (\%)}& \multicolumn{3}{c}{F1-score (\%)}
\\
\hline
& SBP& DBP& MBP& SBP& DBP& MBP
\\
\hline
\multicolumn{7}{c}{Test-I}
\\
\hline
Excluded  & 69.2& 75.0& 75.3& 69.7& 75.5& 75.7\\
Included & \textbf{71.3}& \textbf{76.4}& \textbf{76.2}& \textbf{71.8}& \textbf{76.9}&\textbf{76.7}
\\
\hline
\multicolumn{7}{c}{Test-II}
\\
\hline
Excluded & 83.9& 81.5& 82.1& 87.2& 85.5& 86.5\\
Included & \textbf{85.4}& \textbf{87.3}& \textbf{88.8}& \textbf{88.5}& \textbf{89.7}&\textbf{90.8}
\\
\hline
\end{tabular}
\end{table}

\subsection{Threshold evaluation}
\begin{figure}[!t]
\centerline{\includegraphics[width=0.8\columnwidth]{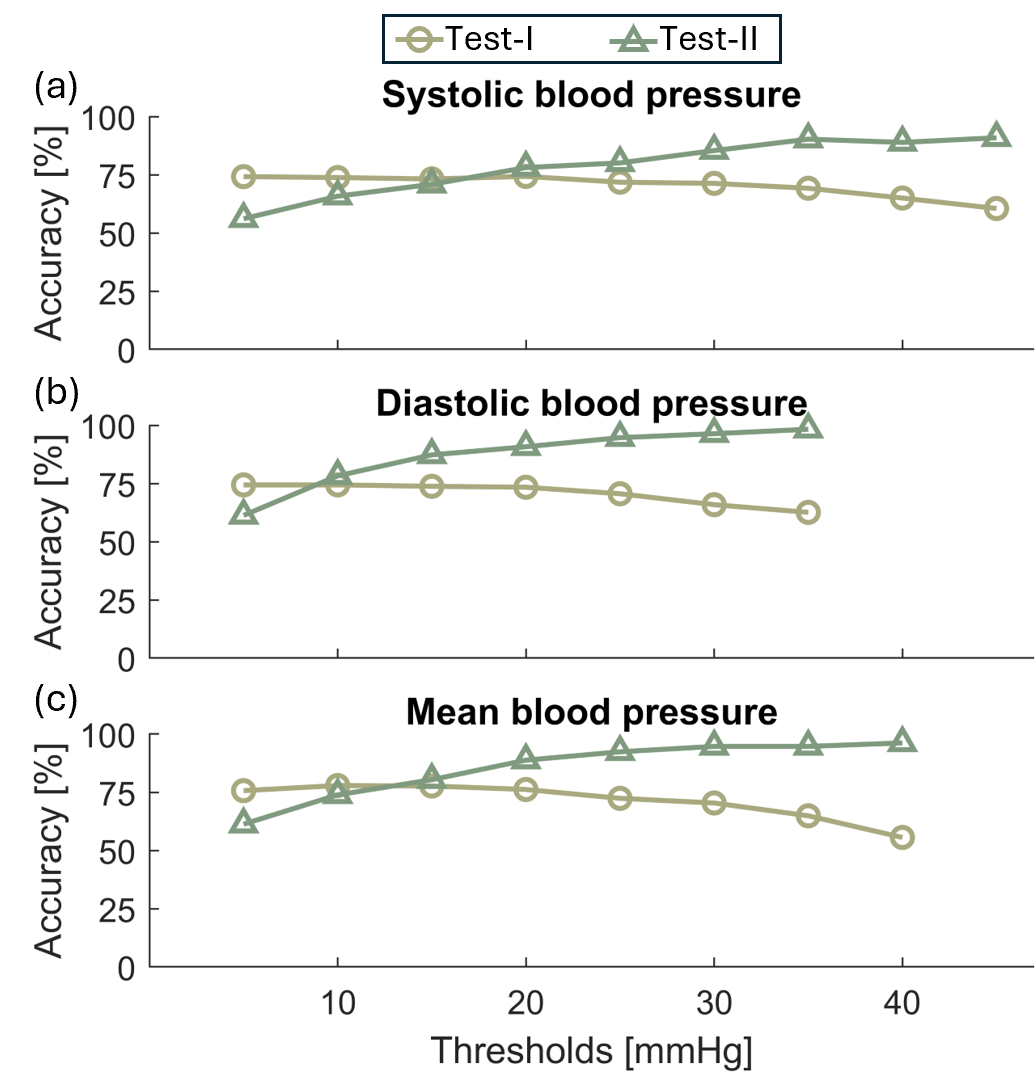}}
\caption{Accuracy of the Encoder model with the PPG-sdPPG-waveform input across different blood pressure (BP) thresholds. Results are for the detected changes in (a) systolic, (b) diastolic, and (c) mean blood pressures evaluated on the Test-I (circles) and Test-II (triangles) datasets}
\label{fig5}
\end{figure}

The accuracy of detection labels for changes in SBP, DBP and MBP produced by the Encoder model with PPG-sdPPG-waveform input decreased with increasing thresholds on the Test-I dataset, while it increased on the Test-II dataset (Fig. \ref{fig5}). On the Test-I dataset,  accuracy peaked at about 75\% with a 5-mmHg threshold, but decreased to around 60\% at higher thresholds (45 mmHg for SBP, 35 mmHg for DBP, and 40 mmHg for MBP). Conversely, the model showed poor classification results at a 5-mmHg threshold on Test-II, with an average accuracy of 60\% for all BP types, while the detection accuracy increased and approached 100\% at the maximum thresholds. It is noteworthy that, since there was insufficient data on changes in DBP and MBP greater than 35 mmHg and 40 mmHg, respectively, the maximum threshold for DBP was tested at 35 mmHg and for MBP at 40 mmHg. The F1-score, varied similarly to detection accuracy across thresholds for all BP types (Supplementary Material Fig. \ref{fig_sup6}).

Supplementary Material Figs. \ref{fig_sup1} to \ref{fig_sup5} compared classified BP changes for all five patients in the Test-II dataset with reference values, under varying threshold settings. The proportion of classified labels that fell within the ‘Stable’ range increased with the increasing threshold, for all BP types. For all threshold values, the model accurately detected the Spike and Dip labels when the reference BP suddenly increased or decreased beyond the thresholds for all BP types, detecting changes in blood pressure earlier the decreasing threshold values. Mislabeled changes often occurred when BP fluctuations were within the set thresholds, but the model could recognize the direction of the fluctuation even if it did not exceed the threshold value.

\section{Discussion}
Our study developed a calibration-free classification model that uses only the PPG signal and an initial BP value to label changes in SBP, DBP, and MBP over hours, thereby achieving the goal of real-time BP monitoring. This project involved a three-category classification task with time-series input data. Initially employing a multi-layer perceptron (MLP) classification accuracies and F1-scores were above 62\% for all BP types across two test datasets (Test-I and Test-II), suggesting a correlation between PPG morphology and changes in BP. More complex models, including convolutional neural network (CNN), residual network (ResNet), and Encoder model based on the Softmax weighting process, improved performance. The Encoder model, in particular, showed significant improvements due to its ability to focus on highly relevant parts of the time-series data. As a result, only this model was further studied to assess the effect of different input types, ablation of initial BP values, and threshold values.

Incorporating second-derivative PPG (sdPPG) waveforms further improved the classification results indicating that sdPPG contains valuable information related to BP changes. This aligns with previous findings that sdPPG is useful for accurately estimating estimating absolute BP \cite{ma_ppg-based_2024} and assessing vascular aging and arterial stiffness \cite{wowern_digital_2015}. Directly inputting waveform data into the first CNN block of the Encoder model proved more effective than manually extracting features and feeding them into each layer of the model through linear layers. The ablation study showed that incorporating initial BP values helps improve detection accuracy, consistent with previous research on absolute BP estimation \cite{mieloszyk_comparison_2022,gao_utransbpnet_2023}.

Contrasting trends in detection accuracy with increasing threshold values between the Test-I and Test-II datasets were attributed to their different data sampling strategies. During the sampling process for the training and validation dataset, patient selection was adjusted based on changes in thresholds to ensure the same volume of data for each classification label. The Test-I dataset used uniform distribution sampling to the training and validation dataset, resulting in a narrow Stable region with small thresholds. Consequently, the narrow Stable region exhibited more consistent or similar patterns of BP changes, making it easier to differentiate from Spike or Dip states, improving classification outcomes. In contrast, the Test-II dataset did not involve selective sampling and utilized the complete set of patient data, resulting in a distribution of BP changes that approximated a normal curve, with most data centred in the middle. When the threshold was increased, enlarging the Stable range, most of the data fell within this expanded Stable range. This shift allowed the same model to achieve better classification performance, as most of the data were classified as Stable.

Compared to the threshold settings of BP changes used in this study, various real applications employed similar thresholds to identify significant BP changes. In clinical settings, ICU patients with acute severe hypertension (SBP/DBP > 180/110 mmHg) were recommended to have their SBP reduced by no more than 25\% within the first hour, aiming for 160/100–110 mmHg over the next 2–6 hours \cite{yu_expert_2023}. Therefore, significant BP changes are defined as at least 20 mmHg for SBP and 10 mmHg for DBP for effective monitoring. Similarly, for acute ischemic stroke patients, a decrease in SBP of more than 26 mmHg within 4 hours after admission is an important indicator for evaluating long-term outcomes \cite{ritter_blood_2009}. Outside clinical settings, a 20-mmHg increase in nighttime SBP has been linked to increased cardiovascular disease (CVD) event risk \cite{kario_nighttime_2020}, highlighting the broader implications of detecting BP changes for preventing CVDs. Overall, the thresholds for BP changes slightly vary across different contexts but generally fall within a 30-mmHg range for SBP, or are dynamically adjusted based on baseline BP to fulfill detection requirements.

The Encoder model could accurately detect BP changes for all BP types, despite occasional lag errors and threshold misjudgements compared to the actual measured BP values. When the BP suddenly dropped below the setting threshold and then returned to the Stable range, the model outputted the Dip state after the BP had returned to Stable. This could be due to the hysteresis effect between PPG and BP \cite{xing_temporal_2023} or measurement errors in PPG \cite{qin_advances_2022}. Threshold misjudgement occurred because the model effectively detected sudden changes in BP but was less accurate in estimating the absolute BP values. As a result, the model outputted Spike or Dip states even when BP values had not actually reached the threshold.

The trained Encoder model detected BP changes across patients without a calibration process involving demographic data, such as age and sex, or PPG data from the test datasets to fine-tune the trained model, departing from previous studies \cite{ma_ppg-based_2023, hong_deep_2021}. This suggests that PPG signals contain BP change information unaffected by demographic factors and, therefore, studies on estimating BP values using PPG can further explore the potential of calibration-free BP estimation models. 

This study has several limitations that need to be addressed in future research. Due to computational limitations, only a portion of patients from the entire VitalDB dataset was used for training, validation, and testing. Although the sampling process was randomized, more accurate results might have been obtained if the entire dataset had been used for the model training. In addition, VitalDB only recorded physiological signals in ICU patients, meaning that most BP changes were induced by medical interventions or the patients' underlying diseases. Therefore, the applicability of the model trained on this dataset to the healthy population requires further investigation.

All models proposed in this study were suitable for time-series classification tasks and have relatively simple structures. It is worth noting that the aim of this study was not to develop a novel model specifically for real-time detection of BP changes; instead, the goal was to use existing models to achieve this purpose. The codes for training and testing all models and generating all datasets are freely available on GitHub [link will be added if the paper is accepted].

\section{Conclusion}
This study developed an Encoder-based model that uses only PPG signals and an initial BP value to continuously and in real-time classify BP changes in ICU patients, without individual calibration. It achieved high accuracy in detecting changes in SBP, DBP, and MBP, demonstrating potential for real-time clinical BP monitoring. The model's simple architecture allows for future investigations of more complex time-series classification models. Testing on broader datasets, including healthy cohorts, is needed to assess wider applicability.

\appendices
\section{}
\label{app:features}
The feature selection for the second-derivative PPG (sdPPG) was conducted using an in-silico dataset comprising 4,374 virtual subjects \cite{charlton_modeling_2019}. The PulseAnalyse algorithm was employed to extract 40 features from the in-silico PPG, first-derivative PPG (dPPG), and sdPPG signals \cite{charlton_assessing_2018}. These features were analyzed for correlation with systolic blood pressure (SBP) to rank their correlation coefficients. The top five correlated features, all derived from sdPPG, were selected for the input analysis. The details of these selected features are presented in Table \ref{tab3}.


\section*{References}

\bibliographystyle{ieeetr}
\bibliography{references}

\renewcommand{\thefigure}{S\arabic{figure}}
\setcounter{figure}{0}

\begin{figure*}[!t]
\centerline{\includegraphics[width=\columnwidth]{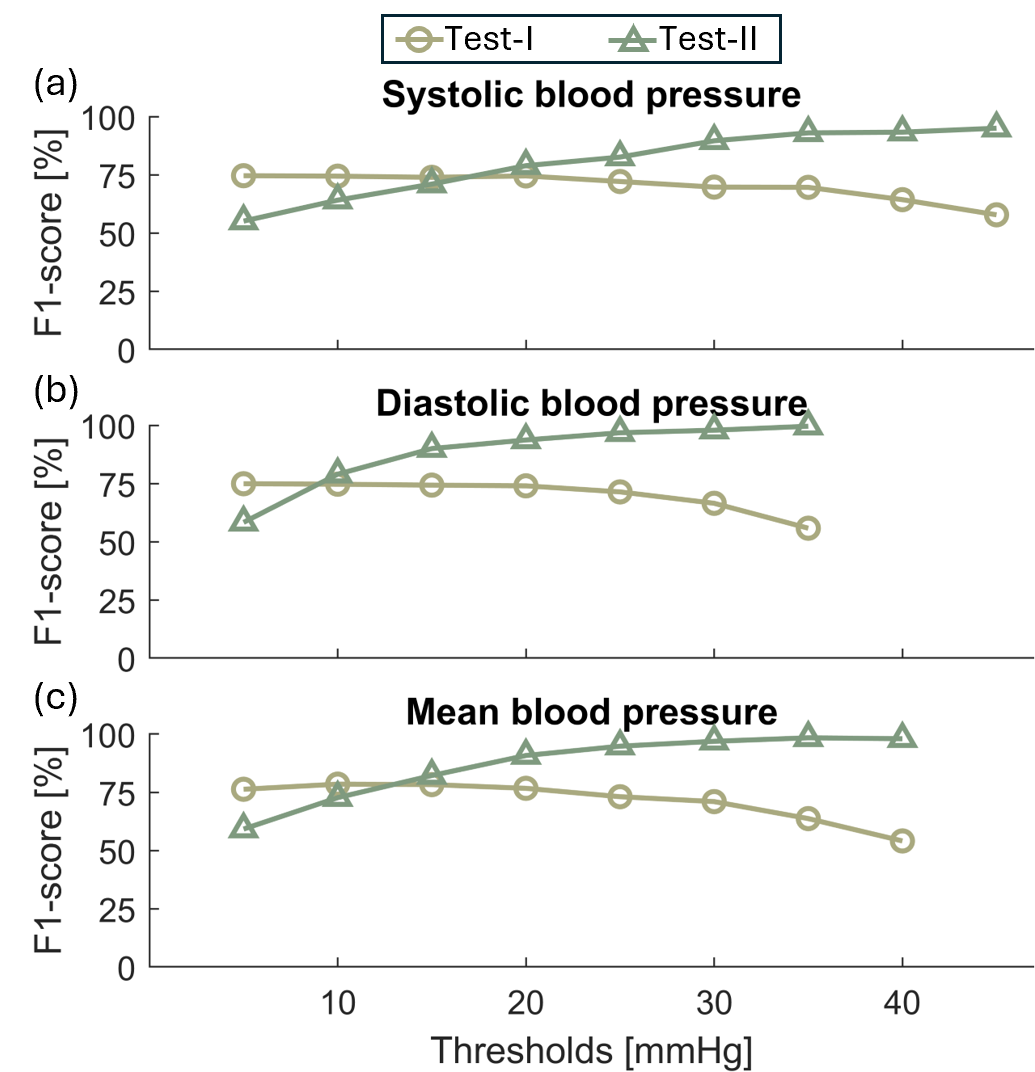}}
\caption{F1-score of the Encoder model with the PPG-sdPPG-waveform input across different blood pressure (BP) thresholds. Results are for the detected changes in (a) systolic, (b) diastolic, and (c) mean blood pressures evaluated on the Test-I (circles) and Test-II (triangles) datasets}
\label{fig_sup6}
\end{figure*}

\begin{figure*}[ht]
    \centering
    \includegraphics[height=\textheight]{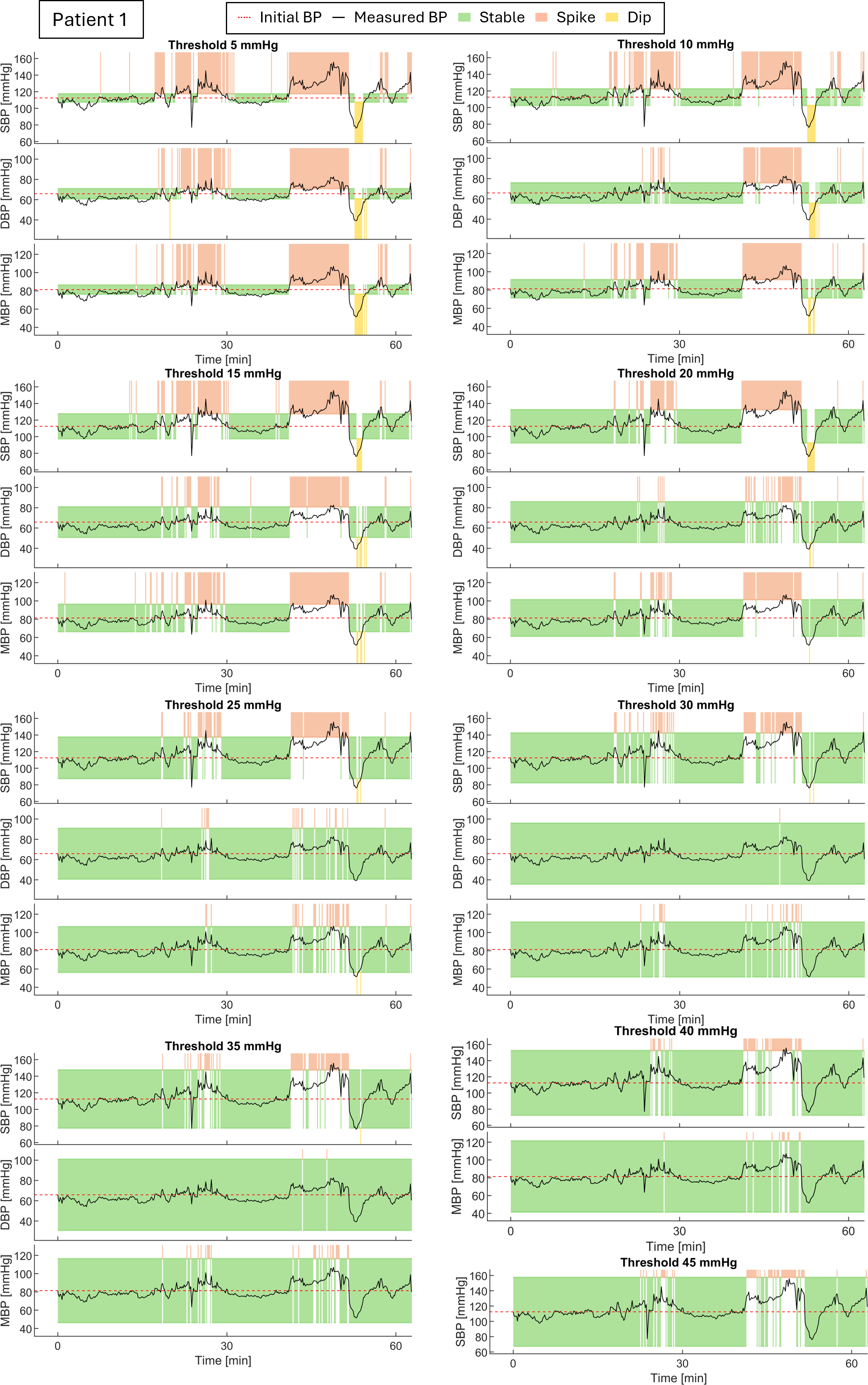} 
    \caption{Label detection results for Patient 1 in the Test-II dataset using the Encoder model with the PPG-sdPPG-waveform input type at various thresholds. For each threshold, reference SBP, DBP and MBP (black lines) are shown together with corresponding detected Spike (red), Stable (green), and Dip (yellow) labels when setting initial pressures at time zero (dashed red lines).}
    \label{fig_sup1}
\end{figure*}

\begin{figure*}[ht]
    \centering
    \includegraphics[height=\textheight]{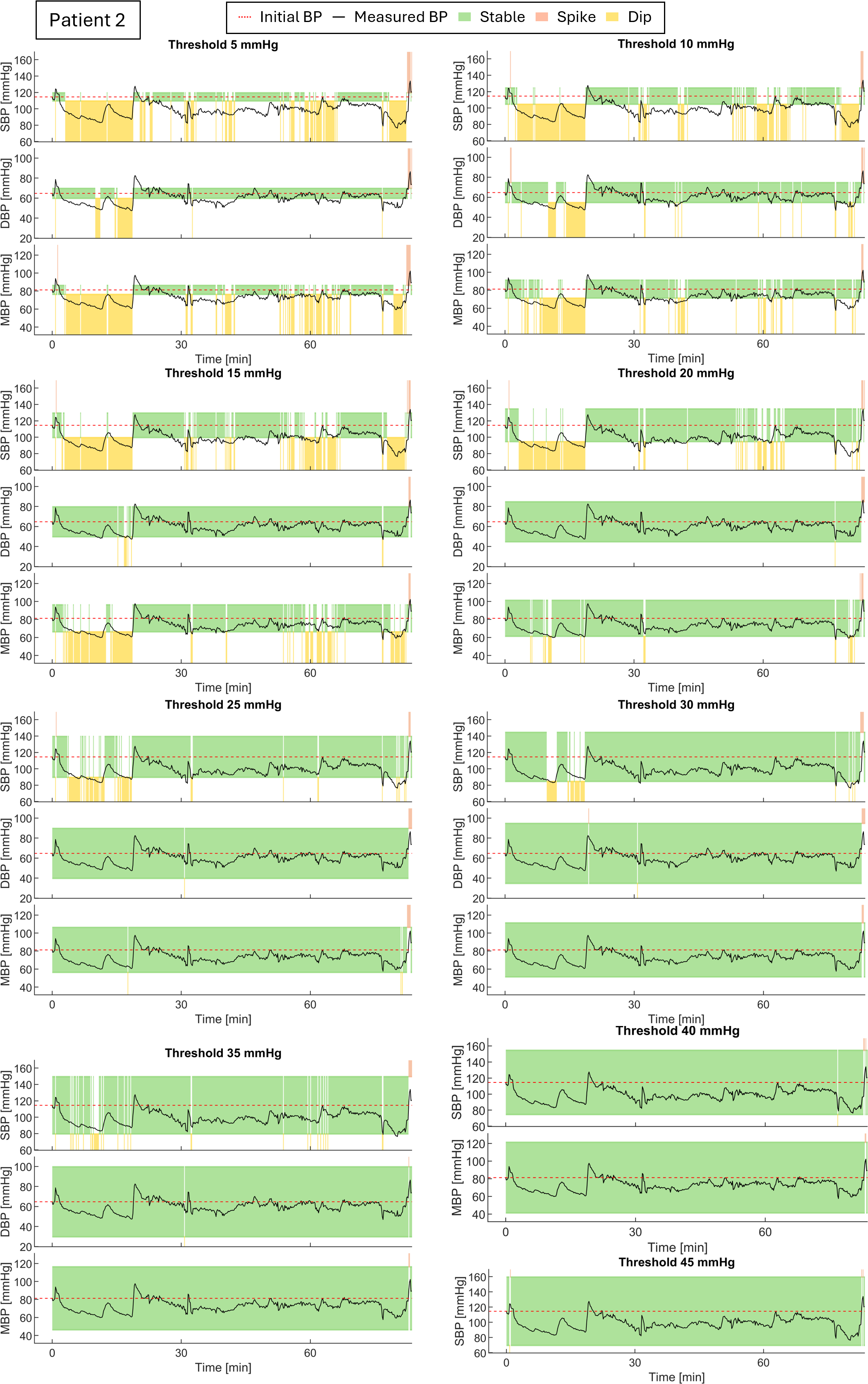} 
    \caption{Label detection results for Patient 2 in the Test-II dataset using the Encoder model with the PPG-sdPPG-waveform input type at various thresholds. For each threshold, reference SBP, DBP and MBP (black lines) are shown together with corresponding detected Spike (red), Stable (green), and Dip (yellow) labels when setting initial pressures at time zero (dashed red lines).}
    \label{fig_sup2}
\end{figure*}

\begin{figure*}[ht]
    \centering
    \includegraphics[height=\textheight]{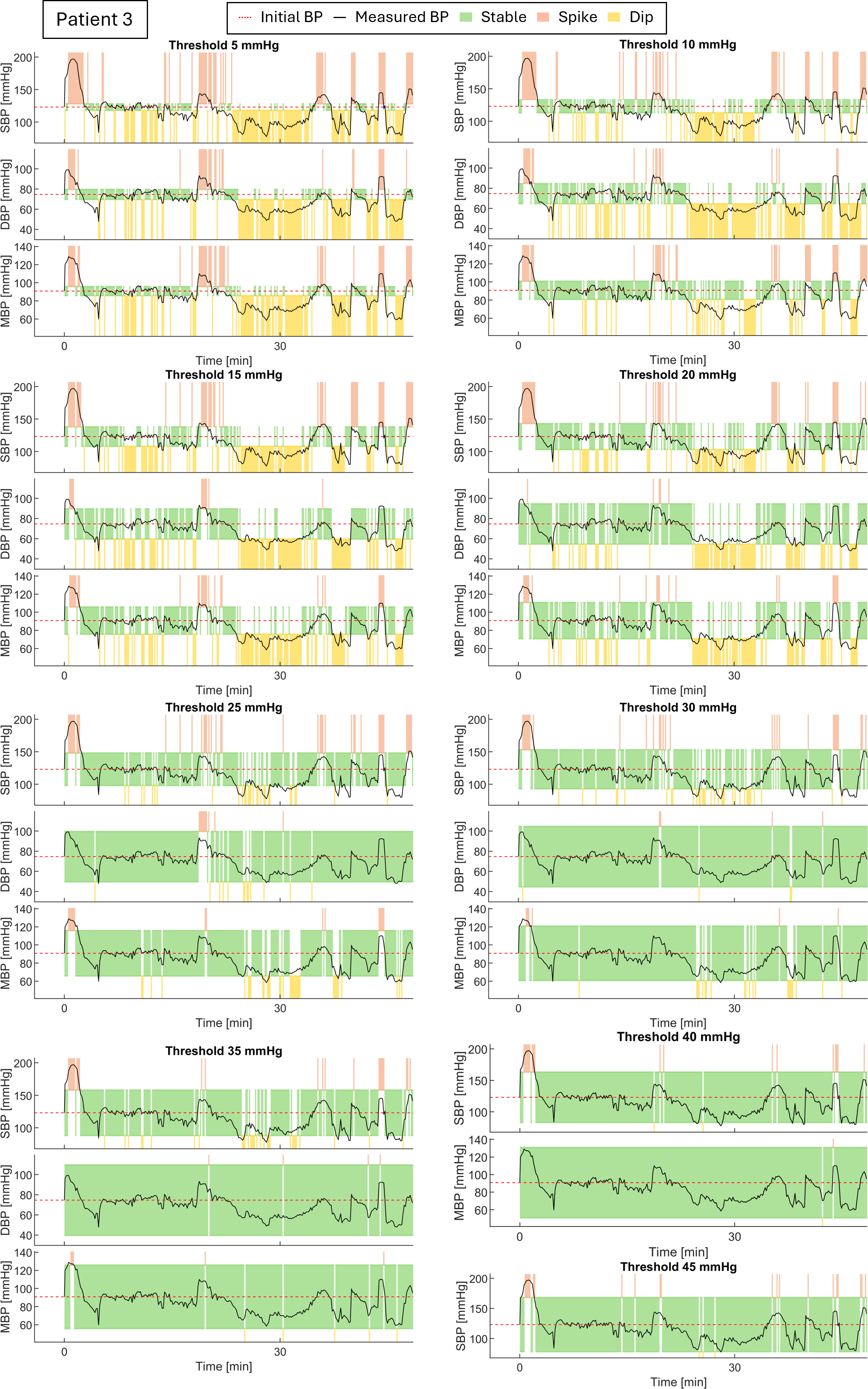} 
    \caption{Label detection results for Patient 3 in the Test-II dataset using the Encoder model with the PPG-sdPPG-waveform input type at various thresholds. For each threshold, reference SBP, DBP and MBP (black lines) are shown together with corresponding detected Spike (red), Stable (green), and Dip (yellow) labels when setting initial pressures at time zero (dashed red lines).}
    \label{fig_sup3}
\end{figure*}

\begin{figure*}[ht]
    \centering
    \includegraphics[height=\textheight]{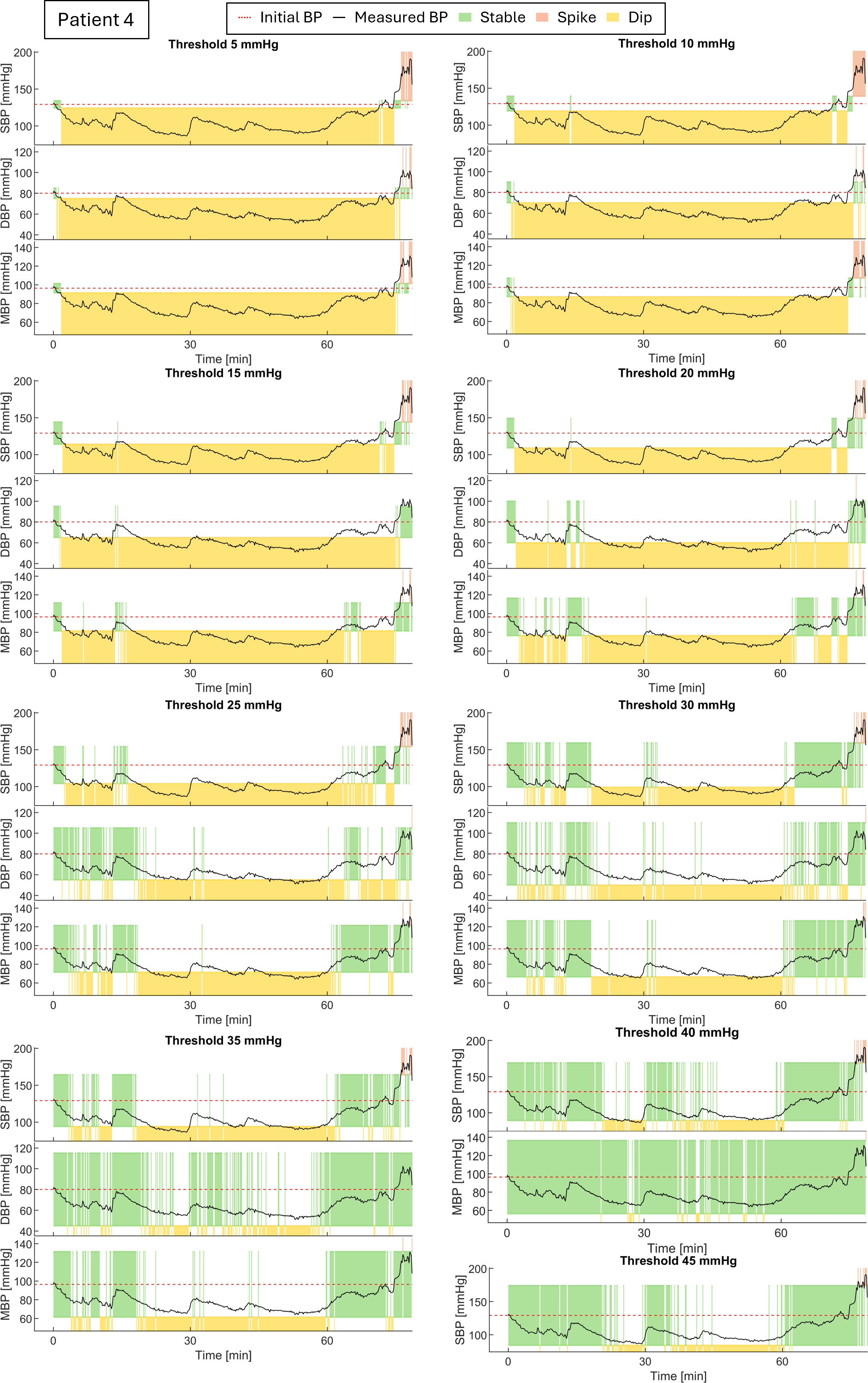} 
    \caption{Label detection results for Patient 4 in the Test-II dataset using the Encoder model with the PPG-sdPPG-waveform input type at various thresholds. For each threshold, reference SBP, DBP and MBP (black lines) are shown together with corresponding detected Spike (red), Stable (green), and Dip (yellow) labels when setting initial pressures at time zero (dashed red lines).}
    \label{fig_sup4}
\end{figure*}

\begin{figure*}[ht]
    \centering
    \includegraphics[height=\textheight]{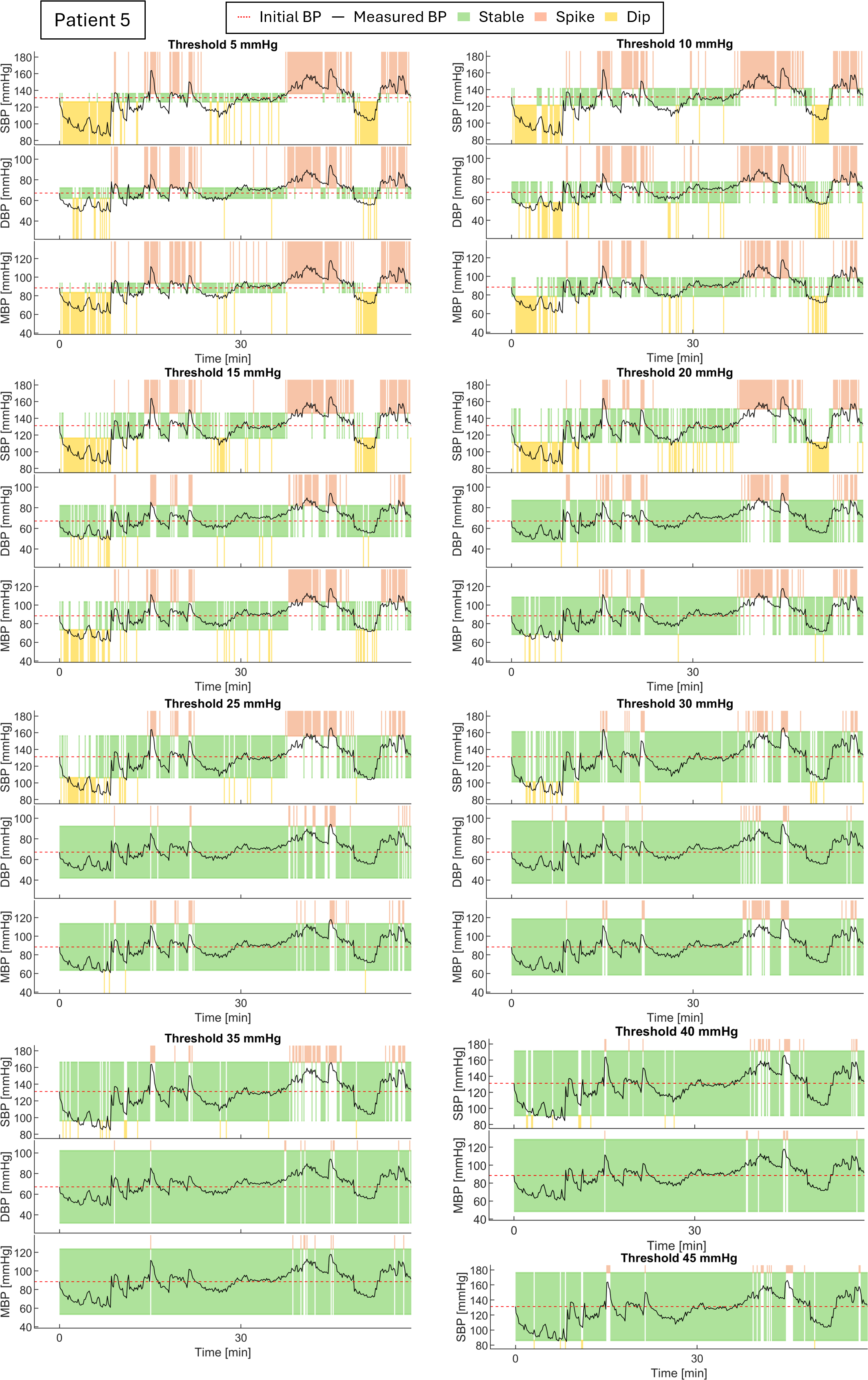} 
    \caption{Label detection results for Patient 5 in the Test-II dataset using the Encoder model with the PPG-sdPPG-waveform input type at various thresholds. For each threshold, reference SBP, DBP and MBP (black lines) are shown together with corresponding detected Spike (red), Stable (green), and Dip (yellow) labels when setting initial pressures at time zero (dashed red lines).}
    \label{fig_sup5}
\end{figure*}

\end{document}